\documentclass{webofc}
\usepackage[varg]{txfonts}   
\begin{document}
\title{Initial-state and final-state effects on hadron production in small collision systems}
%
%

\author{\firstname{Ivan} \lastname{Vitev}\inst{1}\fnsep\thanks{\email{ivitev@lanl.gov}} \and
        \firstname{Weiyao} \lastname{Ke}\inst{1}\fnsep\thanks{\email{weiyaoke@ccnu.edu.cn}} 
             }

\institute{Los Alamos National Laboratory, Theoretical Division, Los Alamos, NM 87545, USA
          }

\abstract{%
Heavy meson production in reactions with nuclei is an active new frontier to understand QCD dynamics and the process of hadronization in nuclear matter. Measurements in various colliding systems at RHIC and LHC, including Pb-Pb, Xe-Xe, O-O, p-Pb, and p-O, enable precision tests of the medium-size, temperature, and mass dependencies of the in-medium parton propagation and shower formation. We employ a coupled DGLAP evolution framework that takes advantage of splitting functions recently obtained in soft-collinear effective theory with Glauber gluons (SCET$_{\rm G}$) and hard thermal loop (HTL) motivated collisional energy loss effects. With jet quenching effects constrained to the nuclear modification factor of charged hadrons in Pb-Pb collisions at 5.02~TeV, we present predictions for light and heavy-meson in Xe-Xe, O-O and p-Pb collisions at the LHC. We find that the nuclear modification scales non-trivially with the quark mass and medium properties. In particular, there can be sizeable collision-induced attenuation of heavy mesons in small systems such as oxygen-oxygen and high-multiplicity p-Pb events. Finally, we analyze the impact of different models of initial-state parton dynamics on the search for QGP signatures in small colliding systems.
}
\maketitle
\section{Introduction}
\label{intro}

Jet quenching is a clear signature of quark-gluon plasma (QGP) formation in nuclear collisions. Energetic quarks and gluons created in  QCD scattering processes undergo multiple interactions with the constituents in the hot QGP medium, leading to parton shower formation in matter.  As a result, the cross sections for hadron and jet production are significantly suppressed relative  to the proton  baseline scaled by the number of binary collisions $\langle T_{AB}\rangle$. This effect is quantified   
by  the nuclear modification factor $R_{AB}$  in A-B reactions (here shown for hadrons)
\begin{eqnarray}
R_{AB} = \frac{dN_{AB\rightarrow h}(p_T)}{\langle T_{AB}\rangle d\sigma_{pp\rightarrow h}(p_T)}.
\end{eqnarray}
In recent years, measurements in small colliding systems such as $d$-Au and $p$-Pb have produced puzzling results. Collective behavior in the pattern of soft particle production that is attributed to QGP evolution in large systems has been observed, suggestive of possible final-state effects. However, clear evidence of jet quenching has not been established, leading to ambiguity about the nature of the medium produced in small systems. Experiments have been designed to provide further insight into this problem, such as the geometry scan at RHIC using $p$-Au, $d$-Au, and ${}^3$He-Au,  and the O-O  collision runs at  RHIC and LHC~\cite{Citron:2018lsq,Brewer:2021kiv}.  The focus of the theoretical analysis we report~is to combine QGP and full cold nuclear matter (CNM) effects and to contrast results in $p(d)$-A and A-A reactions.  This is an important step to establish the baseline without QGP formation and identify possibly different dynamics in symmetric versus asymmetric small systems.  The purpose of these proceedings is to summarize the findings of this recent systematic investigation~\cite{Ke:2022gkq}    
that analyzes the nuclear modification factors in various large and small systems. Calculations are performed for light hadrons and heavy-flavor mesons, including both cold nuclear matter effects and hot QGP final-state effects -- collisional energy loss and medium-induced radiation corrections to the baseline QCD factorization formalism.
By fixing the jet-medium interaction strength in large colliding systems, we further present predictions for light and heavy-flavor productions in $d$-Au, $p$-Pb, O-O, with and without the assumption of the existence of a QGP.

\section{Theoretical formalism}
\label{th}

\begin{figure}[b]
\centering
\sidecaption
\includegraphics[width=7cm,clip]{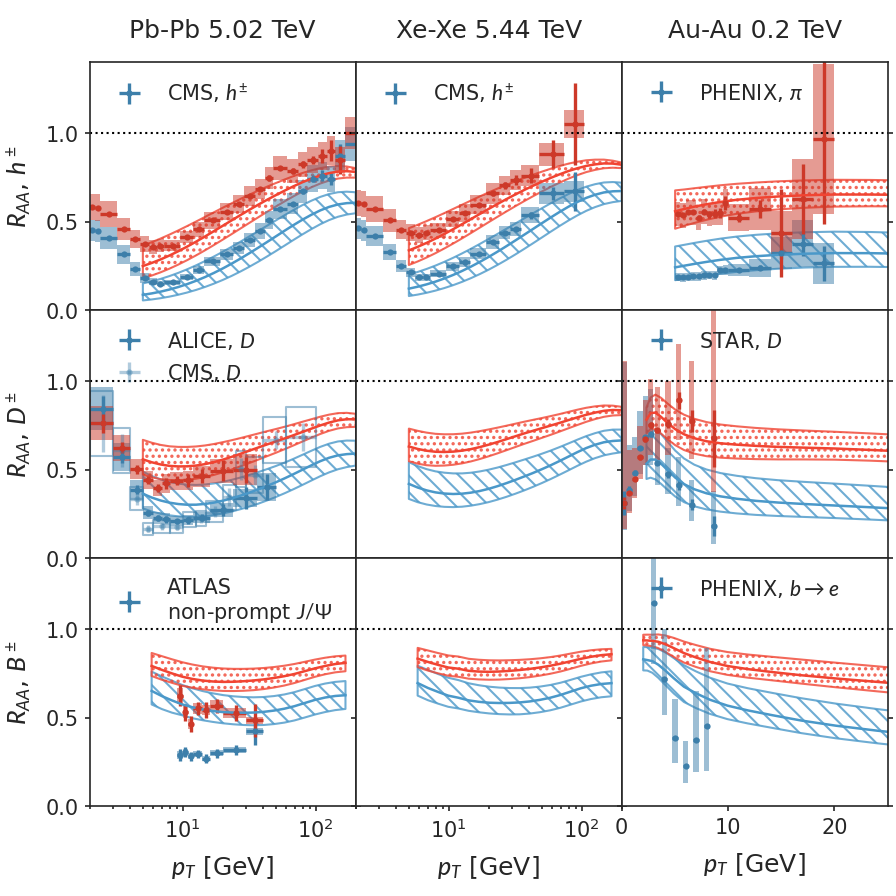}
\caption{From top to bottom rows: nuclear modification factor $R_{AA}$ of light hadron, charm and bottom mesons or their decay products in large colliding systems. 
    From left to right columns: Pb+Pb collisions at $\sqrt{s}=$5.02 TeV for centrality classes 0-10\% and 30-50\%, Xe+Xe collisions at $\sqrt{s}=$5.44 TeV for centrality classes 0-10\% and 30-50\%, and Au+Au collisions at $\sqrt{s}=$200 GeV for centrality classes 0-10\% and 40-50\%. The calculations are compared to measurements by the ALICE, ATLAS, CMS, PHENIX, and STAR experiments. From reference~\cite{Ke:2022gkq}. }
\label{fig-large}       
\end{figure}

Calculations of hadron production in reactions with nuclei  are based on incorporating medium corrections into the peturbative QCD factorization approach. Despite the complicated nature of parton interaction in  nuclear matter, it is possible to model the initial-state effect from QCD interactions, as discussed in~\cite{Vitev:2007ve}. These include the Cronin effect, dynamical shadowing, and cold nuclear matter energy loss. 

In the final state, for the cases where we assume QGP formation, we model the medium evolution using relativistic viscous hydrodynamics with TRENTO initial conditions.  Collisional energy loss is obtained from hard-thermal loop (HTL) calculations. We further evaluate the in-medium branching processes to first order in opacity using SCET$_{\rm G}$, with their soft gluon emission limit interpreted as radiative energy loss. Our analysis shows that collisional processes becomes increasingly important in a small-sized QGP.  For heavy flavor particles, due to the reduced phase space for radiation when $p_T$ is only a few times the heavy quark mass $M$ and the dead cone effect, collisional energy loss is even more important to describe heavy meson suppression at the intermediate $p_T$~\cite{Ke:2022gkq}.
The in-medium parton shower is  implemented in the DGLAP evolution equations, and was recently shown to resum large logarithms of the type  $\ln(E/\mu_D^2L)$ in matter \cite{Ke:2023ixa}.

\section{Phenomenological results}
\label{pheno}

We start by considering the nuclear modification factor $R_{AA}$  in large collision systems for central and mid-peripheral collisions. Results presented in figure~\ref{fig-large} show good agreement between theory and measurements of light and charm hadrons. Tensions still remain for the B meson case (measured through non-prompt $J/\psi$ or non-photonic electron decays). We find that final-state interactions are responsible for the dominant nuclear matter effects in these Pb-Pb,  Xe-Xe,  and Au-Au systems.  

\begin{figure}[b]
\centering
\sidecaption
\includegraphics[width=6.3cm,clip]{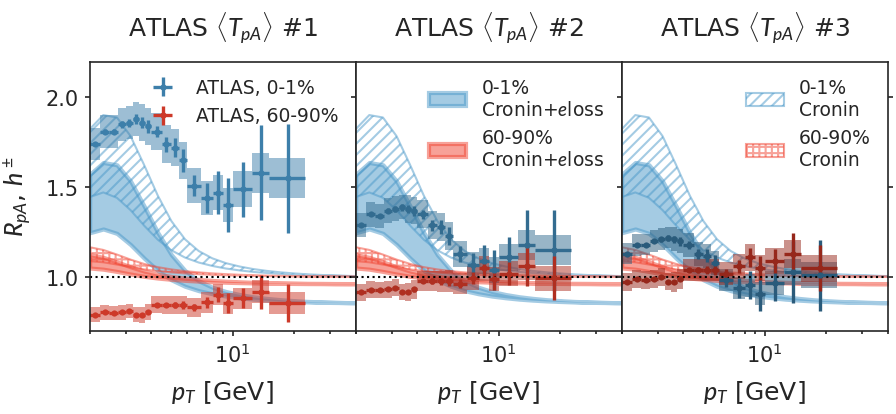}  \ \ 
\includegraphics[width=6.3cm,clip]{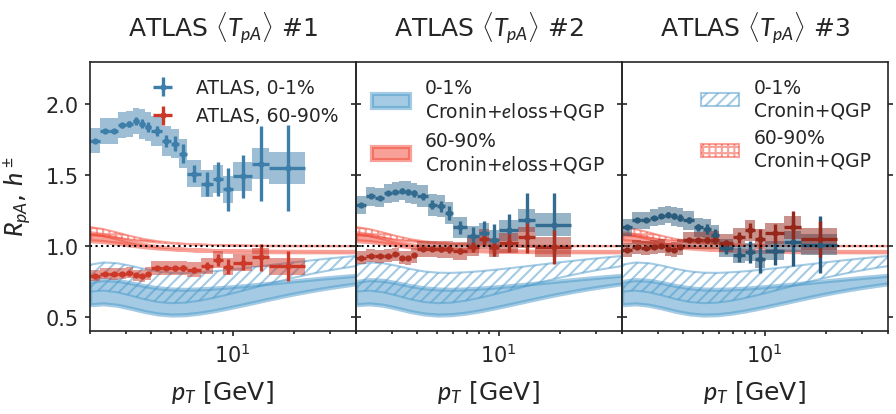}
\caption{Left: nuclear modification factor $R_{p\textrm{Pb}}$ compared to ATLAS data scaled by the overlap functions from three different calculations of nuclear collision geometry. The results for 0-1\% and 60-90\% centralities are shown in blue and red, respectively. The calculations only include cold nuclear matter effects. The shaded bands include dynamical shadowing and Cronin effect, while the filled bands further reflect consideration of CNM energy loss. Right: same as left, but with QGP effect (elastic and radiative) included. From reference~\cite{Ke:2022gkq}.  }
\label{fig-small}       
\end{figure}

Next, we turn to a small asymmetric system, such as $p$-Pb shown in figure~\ref{fig-small}. On the experimental side, the published $R_{pA}$ data are strongly model-dependent and the ATLAS Collaboration obtains the normalization $\langle T_{pA}\rangle$ in the conventional Glauber model, and the improved Glauber-Gribov model with two choices of a parameter that controls the proton fluctuation.  We consider the geometric models \#2 and  \#3 to be much more realistic from the point of view that $R_{pA}\approx 1$ at large $p_T$. Left panels include theory with only CNM effects. Despite the large model-dependent uncertainty, there is little room left for the hot QGP effects in $p$-Pb in the current calculation. In the right panels, the calculations include both cold and hot nuclear effects. QGP effects introduce a strong centrality dependent suppression of $R_{pPb}$ at intermediate and large $p_T$. This is not consistent with mesurements in either scenario.

\begin{figure}[t]
\centering
\sidecaption
\includegraphics[width=7cm,clip]{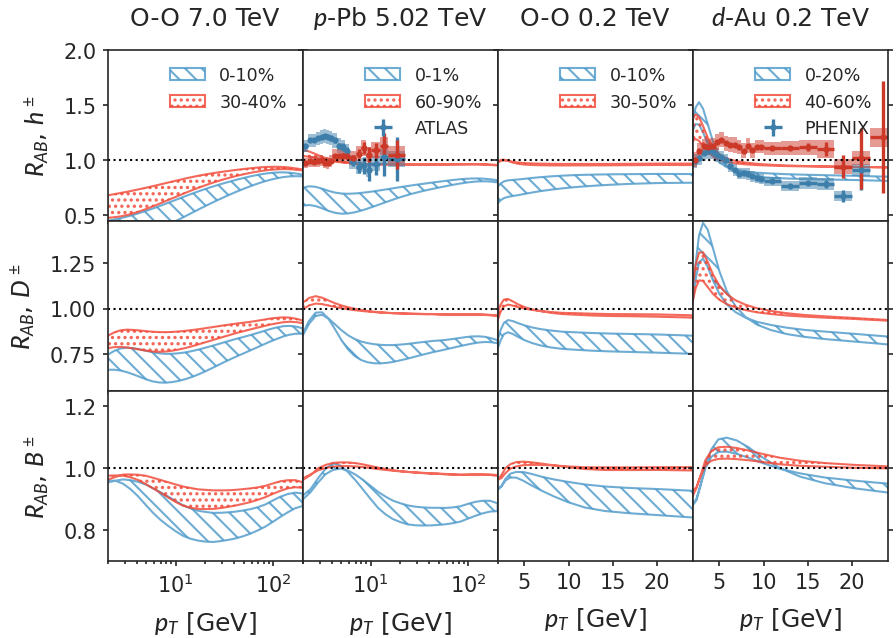}
\caption{From top to bottom rows: nuclear modification factor $R_{AB}$ of light hadron, charm, and bottom mesons in small colliding systems. 
    From left to right columns: O-O collisions at $\sqrt{s}=$7 TeV,  $p$-Pb collisions at $\sqrt{s}=$5.02 TeV, O-O collisions at $\sqrt{s}=$200 GeV, and $d$-Au collisions at $\sqrt{s}=$200 GeV. The $d$-Au data is from the PHENIX Collaboration. These calculations only include both cold and hot nuclear matter effects. From reference~\cite{Ke:2022gkq}. }
\label{fig-OO}       
\end{figure}

Finally, we present predictions for O-O collisions. The advantage of this system is that the correlation between collision geometry and the multiplicity needed to provide unambiguous signatures of nuclear modification in small systems is much more robust than in asymmetric systems. We have performed calculations with only cold nuclear matter effect for O-O, $p$-Pb at the LHC energy  and O-O, $d$-Au at the RHIC energy,  and another set including hot QGP effects. The latter are shown in 
figure~\ref{fig-OO}. If there is no QGP formed in small systems, the CNM effects are small in O-O collisions at both colliding energies. This is in contrast  to asymmetric collisions like $p$-Pb and $d$-Au, where the magnitude of CNM effects depends on the transport properties of cold nuclear matter and phenomenology can only be improved with a better understanding of centrality in $p(d)$-A.
Our study suggests that the O-O system is ideal to search for QGP effects. If a QGP is created in O-O collisions at $\sqrt{s}=7$ TeV, we estimate that hot medium effects can lead to almost 50\% suppression of the charged-particle $R_{AA}$ at $p_T=10$ GeV in central collisions, while bottom mesons can be suppressed by 20\% at $p_T=20$ GeV.
At RHIC, the QGP effect in O-O is predicted to be much weaker from the limited lifetime of the fireball. There is about 20\% suppression in charged particle $R_{AA}$ in central collisions and negligible effects for mid-peripheral collisions.

\section{Conclusions}
\label{concl}

In these proceedings, we reported on a systematic investigation of the modification of light and heavy-flavor hadron production in small and large colliding systems at moderate and high $p_T$. Our goal was to differentiate the impact of cold nuclear matter and hot QGP effects. We found the calculation results to be in a reasonable agreement with the light flavor  and charm mesons suppression  in $A$-$A$ collisions at RHIC and LHC. We further observed that in small colliding systems  CNM effects alone can already explain the basic patterns observed in $p$-Pb  scaled by the improved Glauber-Gribov model. Room for improvement in the description of such systems is still available, as with the CNM transport parameters used here the magnitude of the Cronin enhancement and/or cold nuclear matter energy loss can be overestimated.  In spite of the remaining uncertainties, when we employed a hydrodynamic description of the assumed QGP in $p$-A we found it to lead to quenching of hadron spectra that is inconsistent with the $p$-Pb data. Interestingly,  the same tension was not seen for $d$-Au measurements at RHIC. We finally showed predictions for O-O collisions at RHIC and LHC and found that  CNM effects alone only lead to very small corrections, while the formation of a QGP can suppress charged particle spectra by more than a factor of two at the LHC and by 20\% at RHIC energy. We conclude that if QGP quenching effects are identified in O-O, the enhanced contribution from collisional processes can be tested by simultaneously looking at the flavor dependence of $R_{AA}$. \\

\noindent Acknowledgment: this research is performed in the framework of the HEFTY topical collaboration for nuclear theory. 

%
%

\end{document}